\documentclass[useAMS,usenatbib]{mn2e}
\usepackage{amsmath}
\usepackage{graphicx}
\usepackage{amssymb}

\renewcommand{\v}[1]{\ensuremath{\mathbf{#1}}} 
\newcommand{\uv}[1]{\hat{\ensuremath{\mathbf{#1}}}} 
\newcommand{\abs}[1]{\left| #1 \right|} 
\newcommand{\pd}[2]{\frac{\partial #1}{\partial #2}} 
\newcommand{\grad}[1]{\v{\nabla} #1} 
\renewcommand{\div}[1]{\v{\nabla} \cdot #1} 
\newcommand{\curl}[1]{\v{\nabla} \times #1} 

\title[Stellar Differential Rotation and Coronal Timescales]{Stellar Differential Rotation and Coronal Timescales}

\author[G. P. S. Gibb, M. M. Jardine, D. H. Mackay]{G. P. S. Gibb$^1$\thanks{E-mail: gpsg@st-andrews.ac.uk}, M. M. Jardine$^2$, D. H. Mackay$^1$\\
$^{1}$School of Mathematics and Statistics, University of St Andrews, St Andrews, KY16 9SS\\
$^{2}$School of Physics and Astronomy, University of St Andrews, St Andrews, KY16 9SS}

\begin{document}

\date{Accepted ?. Received ?; in original form ?}

\pagerange{\pageref{firstpage}--\pageref{lastpage}} \pubyear{2014?}

\maketitle

\label{firstpage}

    \begin{abstract} 
    We investigate the timescales of evolution of stellar coronae in response to surface differential rotation and diffusion.
   To quantify this we study both the formation time and lifetime of a magnetic flux rope in a decaying bipolar active region. We apply a magnetic flux transport model to prescribe the evolution of the stellar photospheric field, and use this to drive the evolution of the coronal magnetic field via a magnetofrictional technique. 
Increasing the differential rotation (i.e. decreasing the equator-pole lap time) decreases the flux rope formation time. We find that the formation time is dependent upon the lap time and the surface diffusion timescale through the relation $\tau_\text{Form} \propto \sqrt{\tau_\text{Lap}\tau_\text{Diff}}$.
In contrast, the lifetime of flux ropes are proportional to the lap time ($\tau_\text{Life} \propto \tau_\text{Lap}$). With this, flux ropes on stars with a differential rotation of more than eight times the solar value have a lifetime of less than two days. As a consequence, we propose that features such as solar-like {quiescent prominences may not be easily observable on such stars, as the lifetimes of the flux ropes which host the cool plasma are very short.}  We conclude that such high differential rotation stars may have very dynamical coronae.
    \end{abstract}


\section{Introduction}

The coronae of stars respond dynamically to the emergence and surface flux transport of their star's magnetic field. The surface transport has a number of associated timescales, from the relatively short timescales of flux emergence and differential rotation to the long timescales of stellar cycles. The corona's response to the surface dynamics manifests itself as the star's X-ray luminosity, stellar wind, coronal mass ejections (CMEs) and flares. All of these responses may actively impact on planets orbiting the star. For example, the stellar wind and CMEs apply a torque on the star, causing it to lose angular momentum and spin down \citep{Weber1967,Cameron1989}. On the Sun, the relations between the surface dynamics and the coronal response are well studied 
{(for a review please see \citet{MackayYeates2012})},
however the way in which these relations translate to other stars is not well understood. Previous studies that have considered the stellar coronal responses have found relations between the magnetic flux and the X-ray luminosity \citep{Pevtsov2003}, the magnetic flux and the energy available for driving stellar winds \citep{Schwadron2006} and the relations between stellar flares and CMEs \citep{Aarnio2011,Drake2013}.

On the Sun, prominences are tracers of coronal structure and its dynamics \citep{Mackay2010}. Prominences are found along polarity inversion lines, which separate regions of different magnetic polarity. They are long thin structures of cool dense plasma suspended above the photosphere by magnetic fields. Prominences found within active regions are known as active region prominences. {These short, unstable prominences are associated with solar flares, and tend to be short-lived, with lifetimes of less than two days \citep{Tandberg-Hanssen95,Lites1995,Lites1997,Mackay2010}.} Quiescent prominences are found at the boundaries between active regions, or within decaying active regions. Unlike active region prominences, quiescent prominences are long-lived structures, and can be observed over several solar rotations. Prominences of both types may become unstable and erupt to produce CMEs. {In this study we are concerned with the formation and lifetimes of structures resembling the longer-lived quiescent prominences.}
{The stellar prominences observed to date (also known as slingshot prominences) }are cool dense gas or plasma which has condensed at the tops of long magnetic loops at or around the Keplerian co-rotation radius \citep{Cameron1989,Donati2000}. In contrast, solar prominences are located low down in the corona, with typical heights of at most 100Mm {\citep{priest1982,Tandberg-Hanssen95,Mackay2010}}. Such low lying prominences may be present on other stars but we cannot at present detect them. From now on in the text when we use the term ``prominence'' we refer to a structure resembling a {quiescent} solar prominence.

In order for the cool dense prominence plasma to be supported against gravity, it must be contained within dips of the magnetic field. As such, the downwards weight of the plasma may be balanced by the upwards magnetic tension force of the magnetic field \citep{Kippenhahn57}. Flux ropes -- twisted flux tubes -- have been proposed as magnetic structures that can support prominence plasma, as they contain dips in the magnetic field lines \citep{Kuperus1974,Pneuman1983,Priest1989,vbmartens,Rust1994,Aulanier1998,Gibson2006}. \citet{vbmartens} proposed that a sheared arcade may be transformed into a flux rope due to flux cancellation (see \citet{DeVore2000} for an alternate formation mechanism). For quiescent prominences, one of the sources of shear in the corona is likely to be due to the Sun's differential rotation \citep{vb2000,mackay2006a}. {Additional sources of shear may be from the emergence of sheared field \citep{Pevtsov1995,Leka1996}, the evolution of the large scale properties of active regions \citep{Mackay2011,Gibb2014} or from small scale vortical motions \citep{Antiochos2013,Mackay2014}.}

X-ray and radio observations of cool stars have implied that cool stars exhibit coronae much like the Sun's.
Over the last few decades, observations have shown that cool stars exhibit magnetic fields and dark spots \citep{Strassmeier1996,Donati1997}. Using Zeeman Doppler imaging (ZDI) the distribution of the magnetic fields on such stars may be determined \citep{Semel1989,Brown1991,DonatiBrown1997}. By tracking the stellar spots and magnetic field features obtained from ZDI, the differential rotation profile of these stars may be inferred \citep{DonatiCameron1997,Petit2002}.
Several stars has been found to have lap times - defined as the time for the equator to `lap' the pole - to be much shorter than the Sun's  \citep{Donati2000,Marsden2006,Donati2008,Marsden2011,Waite2011}.
\citet{Barnes2005} found that the lap times decrease with increasing effective temperature of the star, with early G and F type stars having shorter lap times than that of the Sun. They found no correlation between the lap times and the stellar rotation period. \citet{Cameron2007} and \citet{Kuker2011} find relationships between differential rotation and effective temperature that are in qualitative agreement with \citet{Barnes2005}, but have different scaling laws.  \citet{Morin2008} find that M-class stars exhibit solid body rotation. The interpretation of these findings is that the differential rotation rate is inversely proportional to the depth of the convection zone.

The coronal magnetic field of other stars may be modelled in several ways. From ZDI maps of a star's magnetic field, the coronal magnetic field may be extrapolated. The extrapolations typically use the `potential field source surface' method \citep{Altschuler1969}, which assumes that the coronal magnetic field is current-free \citep{Jardine2002,Donati2007,Marsden2011}. 
Whilst good at estimating the global coronal magnetic field, potential fields have the lowest possible magnetic energy for a given boundary magnetic field, and as such cannot be used to determine the energy available to drive flares and CMEs. 
Potential extrapolations may provide a snapshot of the star's global magnetic field, but give no information on the time evolution of the coronal field. ZDI maps for a single star may be obtained at several different epochs, and used to obtain a series of coronal magnetic fields \citep{Donati2008,Fares2009}. It is important to note that these coronal magnetic field extrapolations are produced independently of each other at different times, and cannot represent a continuous time evolution of the coronal field. No information can be obtained about how the differential rotation may shear the star's magnetic field and affect its coronal dynamics. For the Sun, a series of studies has been carried out into the effects of photospheric magnetic flux transport on its large scale coronal field \citep{vb1998,vb2000,mackay2006a,Yeates2008,Yeates2012}.
The evolution of the coronal field of the K0 star AB Dor was modelled by \citet{Pointer2002} using the coronal modelling method of \citet{vb1998}. Further to this, \citet{Cohen2010} ran a magnetohydrodynamical simulation of the corona of AB Dor in order to determine the star's mass and angular momentum loss rates. \citet{Mackay2004} has investigated the photospheric magnetic flux transport on active stars in order to investigate the formation of the observed polar spot caps. This study considered only the evolution of the photospheric magnetic field, and did not investigate the coronal magnetic field evolution.

With evidence that some stars have higher levels of differential rotation than the Sun, it is useful to understand how the enhanced differential rotation affects the dynamics of the stellar corona. In order to address this we consider the formation and stability of flux ropes formed in a simple decaying bipolar active region. We investigate this for different values of differential rotation and surface diffusion. We use a magnetic flux transport model to determine the evolution of the stellar photospheric field. This evolving photospheric field is used to drive the evolution of the coronal magnetic field by applying a magnetofrictional technique. In Section \ref{model} we outline the numerical model we use. In Section \ref{frfinder} we describe our criteria for detecting flux ropes, their formation and eruptions. In Section \ref{results} we provide the results of our study, and finally in Section \ref{discussion} we discuss our results.

\section{The Model} \label{model}
\begin{figure}
\begin{center}
\includegraphics[scale=0.6]{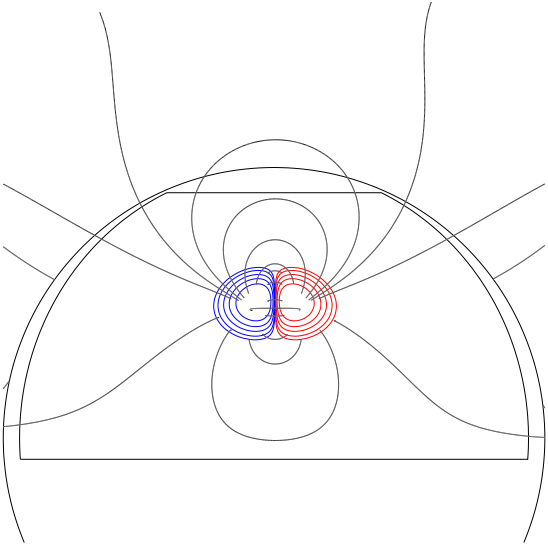}
\caption{Potential initial condition used in this study where red and blue contours represent positive and negative surface flux respectively.  \label{ic_plot}}
\end{center}
\end{figure}
We simulate a portion of a stellar corona using the method of \citet{mackay2006a}. We employ a spherical coordinate system $(r,\theta,\phi)$ where $r$ is the distance from centre of the star, $\theta$ is the co-latitude, related to the latitude, $\lambda$, by $\lambda = 90^\circ - \theta$, and $\phi$ is the azimuthal angle. We simulate the stellar corona between $0^{\circ}$ and $140^{\circ}$ longitude, $-4.5^{\circ}$ and $65^{\circ}$ latitude ($25^\circ$ and $94.5^\circ$ co-latitude), and between radii of $1R_*$ and $2.5R_*$.

We carry out the simulations on an uniformly spaced numerical grid using the variables $(x,y,z)$ defined by:
\begin{align}
x &= \frac{\phi}{\Delta}, \\
y  &= \frac{-\ln \left(\tan \frac{\theta}{2}\right)}{\Delta}, \\
z  &= \frac{\ln \left(\frac{r}{R_*}\right)}{\Delta},
\end{align}
where $\Delta=0.5^\circ$ is the grid spacing. This choice of variables ensures that the horizontal cell size is $h_\phi = h_\theta = r \Delta \sin{\theta}$ and the vertical cell size is $h_z = r\Delta$. We adopt a staggered grid in order to achieve second order accuracy for the computation of derivatives.
We apply a periodic boundary condition on the longitudinal boundaries and a closed boundary condition on the latitudinal boundaries. At the upper ($r=2.5R_*$) boundary we apply an open boundary condition {where the magnetic field, $B$, is assumed to be radial with the electric currents horizontal.} Finally, the lower ($r=R_*$) boundary is specified by the radial photospheric magnetic field as deduced from a 2D surface flux transport model \citep{Sheeley2005}.

\subsection{Surface Flux Transport Model}
In order to model the evolution of the coronal magnetic field with the magnetofrictional method, we require a description of the evolution of the photospheric magnetic field. The photospheric evolution is determined using the flux transport model described in \citet{mackay2006a}. This model assumes the radial photospheric magnetic field, ${B}_r$, is influenced solely through the effects of differential rotation, meridional flows and surface diffusion. The surface diffusion represents the effects of small scale flows such as supergranulation on the large scale field. We express the radial magnetic field at the photosphere by the vector magnetic potentials $A_\theta$ and $A_\phi$ through
\begin{equation}
B_r = \frac{1}{r\sin{\theta}}\left[\pd{}{\theta}(\sin{\theta}A_\phi) -\pd{A_\theta}{\phi}\right].
\end{equation}  
The radial photospheric field is evolved by solving the two dimensional flux transport equation:
\begin{align}
\pd{A_\theta}{t} &= u_\phi B_r - \frac{D}{r\sin{\theta}}\pd{B_r}{\phi} ,\\
\pd{A_\phi}{t} &= -u_\theta B_r + \frac{D}{r}\pd{B_r}{\theta},
\end{align}
where $u_\phi$ is the azimuthal velocity, $u_\theta$ is the meridional flow velocity and $D$ is the photospheric diffusion constant. 

The azimuthal velocity is of the form
\begin{equation}
u_\phi = \Omega(\theta)r\sin{\theta},
\end{equation}
where
\begin{equation} \label{domega scaling}
\Omega(\theta) = K\left(\Omega_0 - d\Omega_\odot \cos^2{\theta}\right) \text{ deg day}^{-1}.
\end{equation}
The term $\Omega (\theta)$ is the angular velocity of rotation relative to the rotation at $30^\circ$ latitude ($60^\circ$ co-latitude). We choose $\Omega_0 = 0.9215\text{ deg day}^{-1}$ and $d\Omega_\odot = 3.65\text{ deg day}^{-1}$ to represent the solar profile. The constant $K$ acts to scale the profile to stars with higher differential rotation rates. Thus we can express the stellar differential rotation rate, $d\Omega_*$, as 
\begin{equation}
d\Omega_* = K d \Omega_\odot.
\end{equation}
Similarly, the quantity $K \Omega_0$ is the angular velocity of $30^\circ$ latitude on the star.

The meridional velocity is prescribed by
\begin{equation}
u_\theta = C\cos\left[\frac{\pi (\theta_{\text{max}}+\theta_{\text{min}} -2\theta)}{2(\theta_{\text{max}}-\theta_{\text{min}})}\right],
\end{equation}
where $C=15 \text{ ms}^{-1}$ is the peak meridional flow velocity of the Sun. The profile is chosen such that the meridional flow vanishes at the latitudinal boundaries $(\theta_{\text{min}},\theta_{\text{max}})$ of the simulation. We adopt the solar meridional flow profile as we have no knowledge of the meridional flow profiles of other stars.

\subsection{Coronal Evolution Model}
We evolve the coronal magnetic field using the ideal induction equation,
\begin{equation}
\pd{\v{A}}{t} = \v{v} \times \v{B},
\end{equation}
where $\v{B} = \curl{\v{A}}$ and
\begin{equation}
\v{v} = \v{v}_{\text{MF}} + \v{v}_\text{out},
\end{equation}
contains contributions from the magnetofrictional velocity ($\v{v}_{\text{MF}}$) and an outflow velocity ($\v{v}_\text{out}$), both of which are described below. Note that we employ the magnetic vector potential, $\v{A}$, as the primary variable in this study as its use in conjunction with a staggered grid ensures the condition  $\div{\v{B}} = 0$ is met.

In the magnetofrictional approach \citep{Yang1986} the equation
of motion of magnetohydrodynamics is modified to include an artificial frictional
term of the form ${\nu}' v$, where ${\nu}'$ is a frictional
coefficient. Under the steady state approximation
and neglecting any external forces, the equation of
motion reduces to:
\begin{equation}
\v{j} \times \v{B} -\nu ' \v{v} = 0,
\end{equation}
where $\v{j} = \curl \v{B}$.
Defining $\nu' = \nu B^2$ the magnetofrictional velocity may then be prescribed by:
\begin{equation} \label{mfvel}
\v{v}_\text{MF} =  \frac{1}{\nu} \frac{\v{j}\times\v{B}}{B^2} .
\end{equation}
The changing photospheric magnetic field -- as specified by the flux transport model -- induces a Lorentz force above the photosphere. The magnetofrictional velocity, which is aligned in the direction of the Lorentz force, acts to advect the coronal field towards a new non-linear force-free equilibrium. 
The changing photospheric field thus drives the evolution of the coronal field through a series of force-free equilibria. 

In addition to the magnetofrictional velocity we also apply a radial outflow velocity of the form
\begin{equation}
\v{v}_\text{out} = v_0 \exp\left(\frac{r-2.5R_{*}}{r_w}\right) \uv{r},
\end{equation}
where $v_0 = 100 \text{ km s}^{-1}$ {and $r_w=0.1R_*$ is the e-folding length over which the radial velocity falls off at the outer boundary.} This outflow velocity is chosen to ensure that the coronal magnetic field at the upper boundary is radial, and also allows any flux ropes that have lifted off from the photosphere to be {completely} removed from the computational box. {Our choice of $r_w$ ensures that the outflow velocity is negligible in the low closed-field corona.} Note that once the field lines become radial near the outer boundary the outflow velocity has no effect on the evolution of the magnetic field.

\subsection{Simulation Set-Up} 
In order to model the photospheric and coronal evolution of the active region, we first must prescribe an initial state. The initial state we choose is a simple bipole whose centre point has latitudinal and longitudinal coordinates of $(\lambda_0,\phi_0) = (30^\circ,70^\circ)$. The half separation between the peaks of positive and negative flux on the photosphere is chosen to be $\rho_0 = 4.5^\circ$. {The bipole's peak flux density at the photosphere is chosen to be $B_0 = -100 \text{ G}$, resulting in a flux of $1.5\times 10^{22}$ Mx -- in agreement with the typical flux of a solar active region.} Finally, the bipole's tilt angle (the angle between the east-west line and the line between the peaks of the positive and negative flux) is chosen to be $\gamma=0^\circ$. We prescribe the radial photospheric field, $B_z(x,y,0)$ according to 
\begin{equation}
B_z(x,y,0) = \frac{B_0 x'}{\rho_0} \exp\left(-\frac{{x'^2}/{2} + y'^2}{2\rho_0^2}\right),
\end{equation}
where
\begin{align}
x' &= (\phi-\phi_0)\cos ({-\gamma}) + (\lambda - \lambda_0)\sin({-\gamma}) \\
y' &= (\lambda-\lambda_0)\cos ({-\gamma}) - (\phi - \phi_0)\sin({-\gamma}).
\end{align}
A potential field in the corona is then calculated from the photospheric field, using the method described by \citet{vb2000}. {The potential field computed assumes that the magnetic field is radial at the upper ($r=2.5R_*$) boundary.}
Figure \ref{ic_plot} displays the initial condition field we use. Note that in all simulations we assume that $R_* =R_\odot$. In Section \ref{tiltangle} we investigate the effects of varying $\gamma$.

\section{Flux Rope Formation and Eruption Criteria}\label{frfinder}
\begin{figure*}
\begin{center}
\includegraphics[scale=1.3]{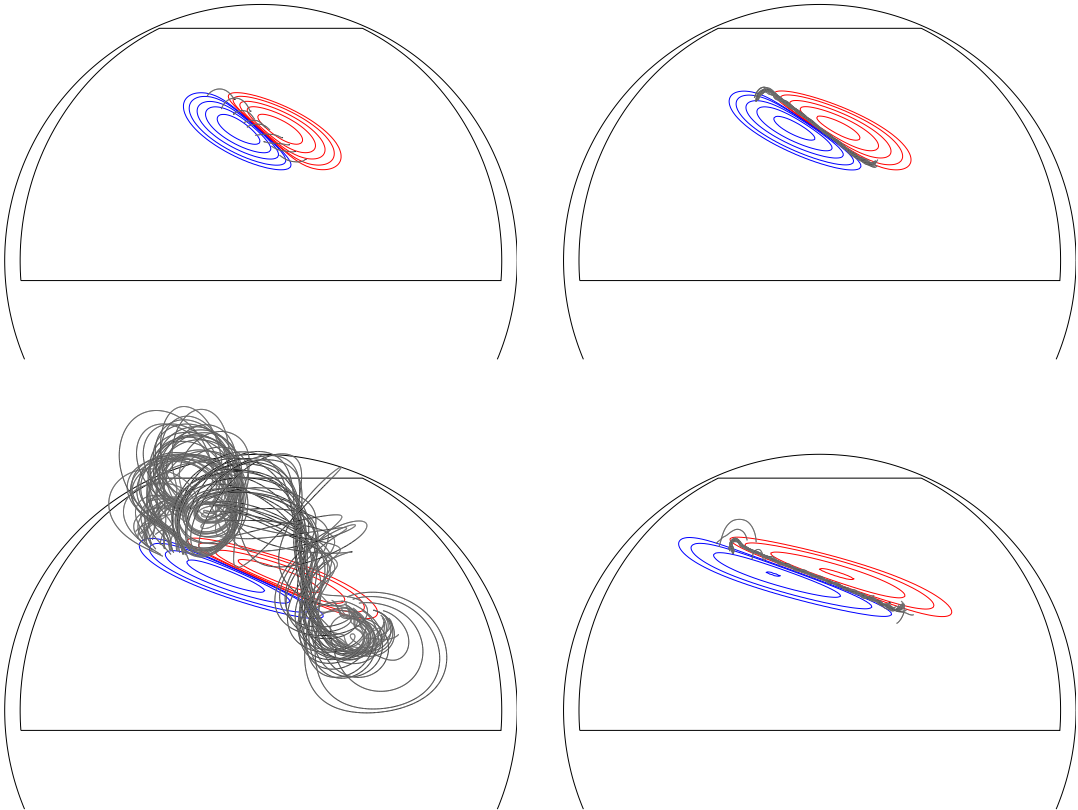}
\caption{Snapshots from the simulation with $d\Omega_*/d\Omega_\odot=3$ and $D=450 \text{ km}^2\text{s}^{-1}$ outlining the sheared arcade transforming into a flux rope, the eruption of the flux rope then the formation of a second flux rope. In each panel the contour levels are the same. \label{selection}}
\end{center}
\end{figure*}
In this section we describe the two methods we use to locate and analyse the flux ropes formed in our simulations. We use two flux rope identification methods to ensure that the quantities determined from our analysis are robust and independent of the nature of the description of the flux rope chosen. Firstly, we will briefly describe the flux rope formation mechanism.

In our simulations the flux ropes are formed above the polarity inversion line within the active region.
The flux ropes form when the arcade field between the two magnetic polarities becomes sheared due to the differential rotation shearing the photospheric flux distribution. Surface diffusion acts to bring the footpoints of the sheared field lines towards each other. The foot points cancel and reconnect, producing a long field line which is strongly aligned with the polarity inversion line. The surface diffusuion continues to bring the foot points of sheared arcades toward each other. Subsequent cancellation and reconnection of these foot points lead to field lines that wrap around the long loop aligned with the polarity inversion line, forming a flux rope. The above formation mechanism is that proposed by \citet{vbmartens}. Several studies, using both synthetic photospheric magnetic field models and observations, have demonstrated that this is a viable formation mechanism for flux ropes on the Sun \citep{Amari1999,vb1999,vb2000,mackay2006a,Gibb2014}. {It is clear from the above description that both shear and flux cancellation are required to form a flux rope. In our simulations, the shear is generated through the effects of differential rotation, whilst the flux cancellation is achieved by the surface diffusion.}

The first method by which we may locate flux ropes is by considering the angle that the horizontal photospheric magnetic field makes with the normal of the polarity inversion line. In order to do this we must first determine the normal vector at every point along the polarity inversion line at the photosphere. This is found by calculating
\begin{equation}
\uv{N}_\text{PIL} = -\frac{\grad B_z}{\abs{\grad B_z}}.
\end{equation}
We then calculate the shear angle, $\theta_s$, by calculating
\begin{equation}
\cos \theta_s = \frac{\v{B}_H \cdot \uv{N}_\text{PIL}}{\abs{\v{B}_H}},
\end{equation}
where $\v{B}_H=(B_x,B_y)$ is the horizontal magnetic field at the photosphere.

By studying the evolution of the shear angle with time we may determine the time when the flux rope forms and the time that it lifts off from the photosphere. For the initial condition, which is a potential field, the shear angle is zero along the entire polarity inversion line. As time progresses in a simulation, the shear angle increases due to the differential rotation shearing the field. In the absence of surface diffusion the shear angle would never reach $90^\circ$ as no field may approach or cross the polarity inversion line. Due to the surface diffusion however, field does reach the polarity inversion line and the resultant flux cancellation and reconnection builds up the flux rope. The signature of a flux rope is a shear angle becoming greater than $90^\circ$ at the photosphere. This is due to the inverse-crossing of the field across the polarity inversion line at the dips of the flux rope's field lines. The existence of a flux rope may thus be inferred by the existence of a shear angle $> 90^\circ$. Through this method the length of the flux rope may be measured by determining the length of the region containing shear angle $> 90^\circ$. {We note that the length as determined by this method underestimates the true length of the flux rope as it only locates where the dips in the flux rope are -- the so called `bald patch' -- and does not detect the extent of the footpoints of the flux rope. Whilst this is the case, the prominence plasma can only be located in the dips of the magnetic field, so in measuring the length of the bald patch, we measure the length of the observable prominence.}

{The ongoing shearing eventually leads to the flux rope becoming unstable and erupting from the simulation. 
{Due to the nature of the lower boundary condition applied, flux cancellation at the polarity inversion line continually occurs. As such, prior to the eruption new field is always being incorporated into the flux rope and it remains in contact with the lower boundary. When the flux rope becomes unstable it lifts off from the lower boundary, leaving behind it a sheared arcade. 
}
The magnetofrictional technique evolves the coronal field through a series of force-free equilibria. It is important to note that the simulation therefore cannot follow the evolution of impulsive events such as an eruption. Within the simulation, after a flux rope has become unstable it may take days for it to fully erupt and be ejected from the computational box. 
After the onset of the eruption, subsequent evolution of the coronal field cannot be trusted as it is no longer in an equilibrium state, and so to the magnetofrictional code cannot follow its evolution correctly. }

The second method with which we can identify a flux rope is by considering the magnetic tension and pressure forces. The Lorentz force may be written as:
\begin{equation}
\v{j} \times \v{B} = \frac{1}{\mu_0}(\v{B}\cdot \grad)\v{B} - \grad\left(\frac{B^2}{2\mu_0}\right),
\end{equation}
where the first term on the right hand side of the equation is the magnetic tension force, and the second term is the magnetic pressure force. At any point along the flux rope's axis, the magnetic pressure force in the plane perpendicular to the axis is directed outwards {from the axis in all directions}. This is because the magnetic field strength is greatest at the centre of the flux rope, and decreases away from the axis. In contrast, the magnetic tension force in the plane perpendicular to the axis is directed inwards {towards the axis in all directions}. This is because the field lines are wrapped around the axis, and thus exert an inwards tension force. These force criteria allow us to determine the points belonging to a flux rope axis in our simulations by looking for locations within the grid where the above criteria are met. 

Using the above method, we may determine the time by which a flux rope has formed in a simulation. We may also study the evolution of the flux rope's length, shape and height with time. When a flux rope becomes unstable, its axis becomes twisted. Thus by studying the evolution of the shape of the axis we may thus gain an indication of the time when the flux rope becomes unstable.
It is also possible to determine the velocity with which the flux rope's axis is moving. This may be calculated using Equation \ref{mfvel}, and a sharp increase in this velocity is also an indication that the flux rope has become unstable.

\section{Results}\label{results}

In this section we consider the effects of varying the differential rotation and surface diffusion coefficient on the formation time and lifetime of flux ropes on solar-like stars. We investigate a range of differential rotation rates between $d\Omega_*/d\Omega_\odot=$ 1--10 (lap times between 98.6 and 9.86 days). This choice approximately covers the range of differential rotation rates greater than the Sun's that have been measured to date. As we have no knowledge of the values of the surface diffusion coefficients on different stars, we also investigate four different surface diffusion constants, namely $D=225,450,900$ and $1800 \text{ km}^2\text{s}^{-1}$ (global surface diffusion timescales ranging from 68--8.5 years). 

Figure \ref{selection} shows a selection of snapshots from a simulation with $d\Omega_*/d\Omega_\odot=3$ (corresponding to a lap time of 32.8 days) and $D=450 \text{ km}^2\text{s}^{-1}$ which highlight the typical evolution of the coronal field in all of the simulations. Firstly, the differential rotation shears the photospheric flux distribution, which results in a sheared arcade field (top left panel of Fig \ref{selection}). Flux cancellation transforms the sheared arcade into a flux rope (top right of Fig \ref{selection}). The continued shearing and cancellation increases the size of the flux rope which eventually leads to the flux rope becoming unstable and lifting off from the photosphere (bottom left panel of Fig \ref{selection}), leaving a sheared arcade. This sheared arcade may form into a second flux rope due to the ongoing differential rotation and flux cancellation (bottom right Fig \ref{selection}). 
In this study we consider two timescales; the timescale for a flux rope to form and the length of time that it may remain stable before eruption - its lifetime.

\subsection{Formation Timescale} \label{formsect}
First, we consider the formation timescale as a function of differential rotation for the four different surface diffusion constants. We define the formation time as the time it takes from the beginning of the simulation for the shear angle to reach $90^\circ$.

\begin{figure}
\begin{center}
\includegraphics[scale=0.5]{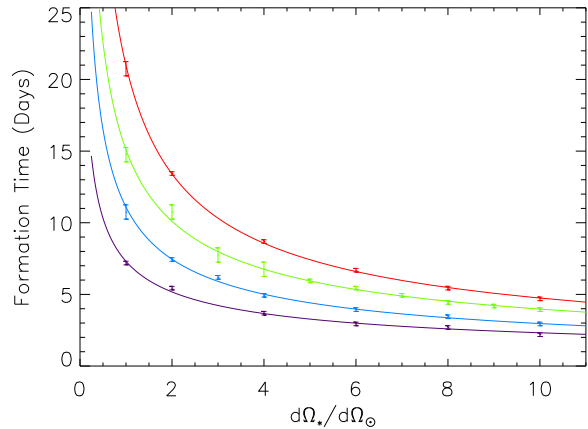}
\caption{Formation timescales as a function of differential rotation for diffusion constants of $250\text{ km}^2\text{s}^{-1}$ (red), $450\text{ km}^2\text{s}^{-1}$ (green), $900\text{ km}^2\text{s}^{-1}$ (blue) and $1800\text{ km}^2\text{s}^{-1}$ (purple). The curves are the power laws fitted to the data as described in Table \ref{table}. \label{formation_plot}}
\end{center}
\end{figure}

Figure \ref{formation_plot} displays the evolution of formation times as a function of differential rotation scaling for various surface diffusion coefficients. It is clear from the plots that for all surface diffusion coefficients investigated, the formation time decreases with increasing differential rotation scaling. Increasing the surface diffusion coefficient has the effect of decreasing the formation times. In order to determine the relation between formation time and differential rotation scaling, we fit the data to a scaling law of the form
\begin{equation} \label{powerlaw}
\tau_\text{Form} = A {\left(\tau_\text{Lap}\right)}^m,
\end{equation}
where $\tau_\text{Form}$ is the formation time, $\tau_\text{Lap}$ is the lap time, $A$ is a scaling constant and $m$ is the power law index for each diffusion constant investigated. Table \ref{table} displays the maximum likelihood estimators for the power law index, $m$, and the scaling constant, $A$, and their $1-\sigma$ confidence intervals for the different surface diffusion constants chosen.  We find that upon increasing the surface diffusion constant by a factor of eight, the power law index decreases by $\approx 25\%$. We thus conclude that the the power law index has a weak dependence on the surface diffusion. We find the mean power law index to be $0.574$ with a standard deviation of $0.06$.

Increasing the surface diffusion constant decreases the scaling constant, $A$.
In order to investigate the dependence of the scaling constant on the surface diffusion, we assume that the power law index, $m$, is independent of the surface diffusion. This assumption is made so that we can directly compare the scaling constants obtained from all four surface diffusion constants investigated. We determine the scaling constant assuming that $m=0.574$ -- the mean value of the power law index found in this study. From the obtained scaling constants, we determine that
\begin{equation}
A \propto D^{-0.44}.
\end{equation}
We therefore tentatively conclude that the formation timescale is approximately determined by
\begin{equation}
\tau_\text{Form} \propto D^{-0.44} {\left(\tau_\text{Lap}\right)}^{0.57}.
\end{equation}
Further to this, if we note that the diffusion time can be expressed as $\tau_\text{Diff} \approx L^2/D$ then we find that approximately,
\begin{equation}
\tau_\text{Form}\propto \sqrt{\tau_\text{Lap} \tau_\text{Diff}}.
\end{equation}
It is very important to note that the scalings determined here are obtained from a range of under one decade in both lap time and surface diffusion constant. The scalings derived must therefore be regarded with caution.

\begin{table}
\begin{center}
\begin{tabular}{c | c c}
$D$ ($\text{ km}^2\text{s}^{-1}$) & $m_{\text{ML}}$ & $A_\text{ML}$ \\ \hline
$225$ & $0.646 \pm 0.009$ & $1.08 \pm 0.01$\\
$450$ & $0.580 \pm 0.015$ & $1.05 \pm 0.01$\\
$900$ & $0.573 \pm 0.017$ & $0.80 \pm 0.01$\\
$1800$& $0.498 \pm 0.011$ & $0.74 \pm 0.01$
\end{tabular}
\end{center}
\caption{The maximum likelihood estimates of the power law index, $m$, and the scaling constant, $A$, from equation \ref{powerlaw} and their $1-\sigma$ confidence intervals for different surface diffusion constants. \label{table}}
\end{table}

\subsection{Lifetime}\label{lifesect}
\begin{center}
\begin{figure}
\includegraphics[scale=0.5]{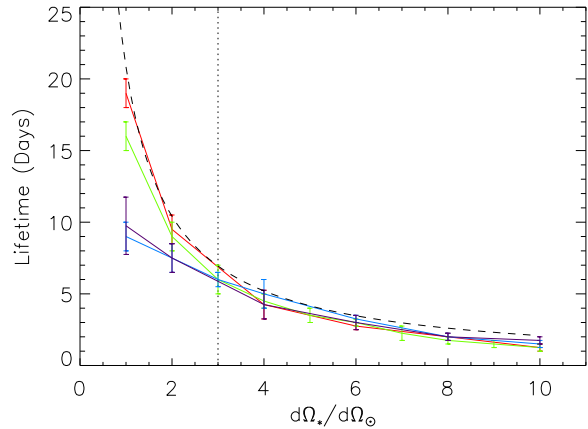}
\caption{Flux rope lifetime as a function differential rotation for diffusion constants of $250\text{ km}^2\text{s}^{-1}$ (red), $450\text{ km}^2\text{s}^{-1}$ (green), $900\text{ km}^2\text{s}^{-1}$ (blue) and $1800\text{ km}^2\text{s}^{-1}$ (purple). The dashed line is the shear timescale (Equation \ref{shearts}). \label{lifetime}}
\end{figure}
\end{center}

Another important timescale to investigate is the lifetime of a flux rope. We define this as the length of time between its formation and the onset of its eruption. As is discussed in Section \ref{frfinder}, there are several methods by which we can define the time of the onset of the eruption. The times derived from each method agree with each other to within two days for low differential rotation stars, and within 0.5 days for high differential rotation stars. Figure \ref{lifetime} displays the evolution of the flux rope's lifetime as a function of differential rotation scaling for the four surface diffusion coefficients investigated. From the plots, it can be seen that for all surface diffusion coefficients investigated the lifetime is inversely proportional to the differential rotation scaling. We find that for stars with differential rotation rates greater than approximately three times the solar value (lap times less than 32 days - highlighted by a vertical dotted line in Fig \ref{lifetime}) the lifetime is independent of the surface diffusion. This may be interpreted as the lifetime being solely dependent upon the shearing caused by the differential rotation. In order to demonstrate this, we may consider the shear timescale. We define this as the time required to build up a shear angle of $45^\circ$ from an initial shear of $0^\circ$ at $30^\circ$ latitude, and find it to be
\begin{equation}
\tau_\text{Shear} = 20.83\frac{d\Omega_\odot}{d\Omega_*} \text{ days.} \label{shearts}
\end{equation}
The shear timescale is represented in Figure \ref{lifetime} by the black dashed curve. It can be seen that the curves for all four diffusion constants investigated are situated below the shear timescale's curve, but generally follow it. We attribute this discrepancy between the lifetime and the shear timescale to the presence of surface diffusion slightly lowering the lifetime.
For differential rotation scalings lower than three (lap times greater than 32 days) we find that the lifetime is dependent on the surface diffusion. In this regime, higher surface diffusion decreases the lifetime. Taking the length scale for diffusion to be three grid cells (the minimum diameter a flux rope must possess to be resolved in the simulation) at $30^\circ$ latitude the diffusion timescale is
\begin{equation}
\tau_\text{Diff}  \approx \frac{L^2}{D} = \left(\frac{450 \text{ km}^2\text{s}^{-1}}{D}\right)6.34 \text{ days.}
\end{equation}
For $d\Omega_*/d\Omega_\odot < 3$ the diffusion timescale for larger surface diffusion constants is much shorter than the shear timescale. We interpret the decrease in the lifetime in this regime to be due to the stronger surface diffusion acting to weaken the arcade above the flux rope, reducing its ability to counter the upwards force from the flux rope with its downwards tension force.

We now summarise the above findings. The lifetime of the flux rope is proportional to the shear timescale, which itself is proportional to the lap time.  Thus
\begin{equation}
\tau_\text{Life} \approx \tau_\text{Shear} \propto \tau_\text{Lap}.
\end{equation}
 The above relation holds unless $\tau_\text{Shear} \gg \tau_\text{Diff}$, whereby the enhanced surface diffusion decreases the lifetime such that
\begin{equation}
\tau_\text{Life} < \tau_\text{Shear}. 
\end{equation}

\subsection{Tilt Angle}\label{tiltangle}
\begin{figure}
\begin{center}
\includegraphics[scale=0.5]{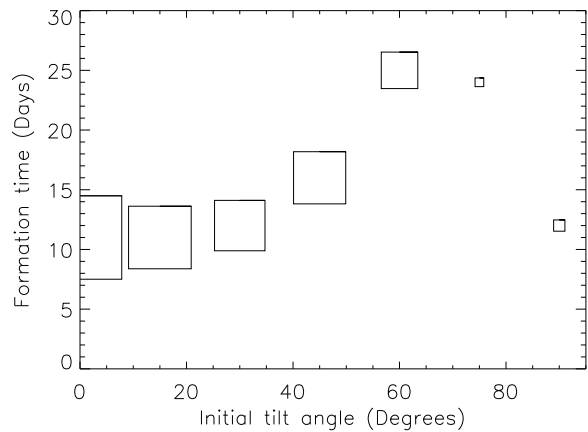}
\caption{Evolution of flux rope formation timescale as a function of initial bipole tilt angle. The symbol size is proportional to the maximum length of the flux rope formed, with the largest and smallest symbols corresponding to flux rope lengths of $48^\circ$ and $5^\circ$ respectively. \label{tilt}}
\end{center}
\end{figure}

\begin{figure}
\begin{center}
\includegraphics[scale=0.6]{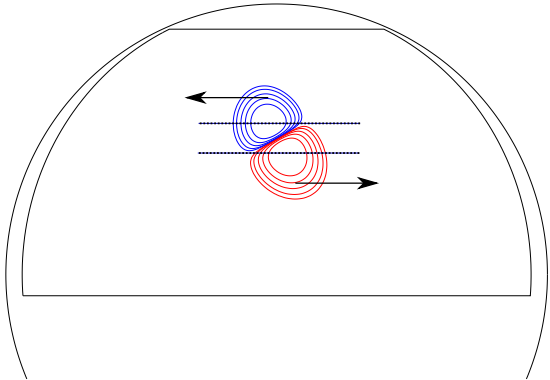}
\caption{Initial condition bipole with tilt angle of $60^\circ$. The region between the dotted lines indicates where a flux rope will form at a later time. The arrows denote the direction the differential rotation advects the surface flux relative to the centre of the bipole. \label{tilt_bipole}}
\end{center}
\end{figure}
 
In the previous subsections of Section \ref{results} we have considered bipoles with initial tilt angles of $0^\circ$. We now investigate the effect that changing the initial tilt angle has on the formation of flux ropes. To achieve this, we ran a set of simulations with $d\Omega_*/d\Omega_\odot=2$ and $D=450\text{ km}^2\text{s}^{-1}$ but varying the tilt angle between $0^\circ$ and $+90^\circ$. The range of tilt angles chosen are consistent with Joy's law, i.e. the leading polarity is closer to the equator than the following polarity. Figure \ref{tilt} displays the evolution of the formation time as a function of tilt angle as determined from the simulations. Also included in Figure \ref{tilt} is an indication of the length of the flux rope formed, represented by the size of the plot symbols used. For tilt angles less than approximately $30^\circ$ the tilt angle has little effect on the formation time, save for a slight increase with increasing tilt angle. The length of the flux rope is found to slightly decrease as the tilt angle is increased. 
For tilt angles greater than $30^\circ$ the formation time increases sharply and the flux rope's length decreases slightly with increasing tilt angle until a tilt angle of approximately $60$-$75^\circ$. 
These effects are caused by the initial bipole tilt angle effectively shortening the length of region where flux cancellation may occur to form the flux rope (region between dotted lines in Figure \ref{tilt_bipole}), as the differential rotation acts to draw the northern edge of the negative polarity region, and the southern edge of the positive polarity region away from the polarity inversion line (arrows in Figure \ref{tilt_bipole}). This results in a flux rope that forms more slowly as diffusion is less able to bring opposite polarity fields together for cancellation. 
For tilt angles above $60$-$75^\circ$ and up to $90^\circ$ the formation time decreases and approaches the formation time for a bipole with initial tilt angle of $0^\circ$. The length of the flux rope remains short, however. This behaviour is different to the behaviour exhibited for tilt angles of less than $60$-$75^\circ$. We attribute this to the tilt angle being sufficiently large that the differential rotation acts to slide the two polarities past each other, resulting in a more efficient shearing of the field. This efficient shearing allows the flux rope to form relatively quickly, however the resultant flux rope is short as {the polarities sliding past each other shorten the length of the region where the flux cancellation, and hence flux rope formation may occur.}
The flux rope formation mechanism for high initial tilt angles is somewhat different to the formation mechanism for tilt angles less than $60$-$75^\circ$ as the shear is driven by the polarities sliding past each other, rather than deformation of the active region by the differential rotation.

\section{Discussion and Conclusions} \label{discussion}

In this study we have considered the effects of differential rotation and surface diffusion on the formation and stability of flux ropes formed in a decaying active region. In order to do this we ran a series of simulations with different surface diffusion coefficients and differential rotation scalings. The simulations consisted of a surface flux transport model to prescribe the evolution of the photospheric magnetic field, coupled with magnetofrictional technique to determine the evolution of the coronal magnetic field due to the evolving photospheric field. 

We found that the formation timescale of a flux rope is approximately proportional to the geometric mean of the equator-pole lap time and the surface diffusion timescale.
The lifetimes of the flux ropes are strongly dependent upon the shearing of the coronal field due to differential rotation. We find that the lifetimes are approximately equal to the shear timescale (Eqn \ref{shearts}), unless the diffusion timescale is much shorter than the shear timescale, whereby the lifetime is shorter than the shear timescale.
We interpret this shortened lifetime as being due to the enhanced diffusion weakening the arcade field that holds down the flux rope below it.
Flux ropes formed from active regions with tilt angles ranging from $0$-$30^\circ$ have similar formation times and lengths. For tilt angles above this the lengths of the flux ropes decrease with increasing tilt angle. For tilt angles between $30^\circ$ and $60$-$75^\circ$ the formation times increase due to a decreased efficiency of diffusion bringing opposite polarity field in to be cancelled. Between $60$-$75^\circ$ and $90^\circ$ the formation time decreases with increasing tilt angle. This is because the increasing tilt results in a more east-west aligned polarity inversion line, which maximises the efficiency of differential rotation to shear the field across it.

\begin{figure}
\includegraphics[scale=0.5]{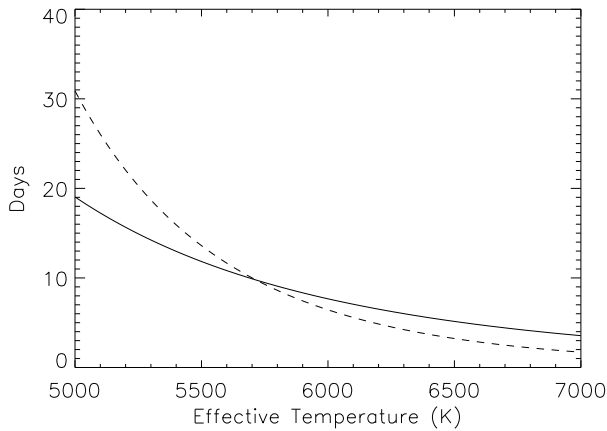}
\caption{Formation time (solid line) and lifetime (dashed line) as a function of stellar effective temperature for a surface diffusion coefficient of $450\text{ km}^2\text{s}^{-1}$.\label{temp}}
\end{figure}

Using the results of \citet{Cameron2007}, who find that the differential rotation of a star is proportional to its effective temperature according to the power law
\begin{equation}
d\Omega_* = 3.03{\left(\frac{T_\text{eff}}{5130\text{ K}}\right)}^{8.6}\text{ deg day}^{-1},
\end{equation}
we may express our results from Sections \ref{formsect} and \ref{lifesect} in terms of the stellar effective temperature. Figure \ref{temp} displays the formation time and lifetimes of flux ropes as a function of stellar effective temperature, with $D= 450\text{ km}^2\text{s}^{-1}$. It can be seen that by increasing the stellar effective temperature from 5000 K to 7000 K the formation time for the flux rope decreases by a factor of $\sim 5$ and the lifetime decreases by a factor of $\sim 18$. This strongly implies that as we move up the main sequence, the evolution timescales of stellar coronae decrease dramatically.

The lifetimes of flux ropes on stars with high differential rotation are considerably shorter than on the Sun. For stars with differential rotation rates greater than four times the solar value, the lifetimes are less than five days. Similarly, for stars with differential rotation greater than eight times the solar value, the lifetime is found to be two days or fewer.
{For such high differential rotation stars, where the lifetime of flux ropes is likely to be less than a few days, we propose that prominences are unlikely to be observed as they are only present on the star for a very short period of time. Whilst the flux rope structure exists for this long, the dips must be populated with prominence plasma in order for the prominence to be visible. Several mechanisms have been put forward to explain how cool plasma comes to be located in the magnetic dips of quiescent prominences. 
Plasma may be injected into the flux rope by reconnection at its footpoints in the chromosphere, forcing cool plasma up into the prominence \citep{Wang1999,Chae2001}. The plasma may also accumulate in the dips of the magnetic field by an evaporation-condensation mechanism. In this scenario heating at the footpoints of coronal loops causes chromospheric material to be evaporated into the loop and heated to coronal temperatures. If the loop is sufficiently long (such as those in flux ropes) this plasma then may condense at the centre of the loop \citep{Serio1981,Mok1990,Antiochos1991,Dahlburg1998} and cool down to chromospheric temperatures \citep{Hood1988}. 
For both of these mechanisms, it is clear that once a flux rope has formed a finite amount of time is required for its magnetic dips to be populated with a sufficient amount of cool plasma for the prominence to be visible. Due to the time required to fill the flux rope with prominence plasma, the prominence may well be present on the star for a shorter period of time than the flux rope's lifetime. Therefore the lifetime we calculate is the maximum amount of time that the prominence may be visible for.
}

On stars with high differential rotation we find that the formation times and lifetimes of flux ropes are significantly shorter than on lower differential rotation stars such as the Sun. We propose that such high differential rotation stars will have far more dynamic coronae, with magnetic structures evolving on much shorter timescales. In each simulation, a series of flux ropes were formed then ejected. On high differential rotation stars, the frequency of eruptions thus may be higher than on low differential rotation stars. An increase in the eruption frequency could result in an increased mass and angular momentum loss from the star.

It is important to note that in the present study we have modelled the decay and shearing of a single, isolated, bipolar active region. No external coronal fields have been included, such as those from other active regions or polar field. Polar field may play a very important role on such stars, as many ZDI observations of stars show polar spots and strong fields \citep{DonatiCameron1997,Donati1999,Donati2003}. The interaction of the active region's magnetic field with an external coronal field may have a significant effect on the formation and stability of the flux rope. Addressing this issue lies outwith the scope of this study.

{
It has long been known that on the Sun active regions tend to possess shear even at the time of emergence \citep{Leka1996}, with active regions in the northern/southern hemisphere generally containing negative/positive magnetic helicity \citep{Pevtsov1995}. In the present study, however, we use a potential field initial condition which possesses no shear. In our simulations, it will therefore take longer form a flux rope from the potential bipole than for the case with an initially sheared bipole. Whilst the flux rope formation time will be decreased for an initially sheared bipole compared to a potential bipole, we believe that the scaling found in this paper will remain the same, namely $\tau_\text{Form} \propto \sqrt{\tau_\text{Lap} \tau_\text{Diff}}$. The lifetime, which is determined by the amount of shear being applied to the flux rope by the differential rotation, should remain unchanged. In a future study, we will address the effects of adding a shear to the initial condition field.
}

We finally summarise the main findings of this paper:
\begin{itemize}
\item We find the formation time of a flux rope scales with the differential rotation lap time and surface diffusion timescale as $\tau_\text{Form} \propto \sqrt{\tau_\text{Lap} \tau_\text{Diff}}$
\item The lifetime of a flux rope scales with the shearing timescale as $\tau_\text{Life} \approx \tau_\text{Shear} \propto \tau_\text{Lap}$
\item {For stars with very high differential rotation the lifetime of flux ropes becomes increasingly short. We propose that prominences may be difficult to observe on such stars as they will only be present for a short time.}
\end{itemize}

\section*{Acknowledgements}
G.P.S.G. would like to thank the STFC for financial support. D.H.M. would like to thank the STFC and the Leverhulme Trust for financial support. Simulations were carried out on a STFC/SRIF funded UKMHD cluster at St Andrews.

\bibliographystyle{mn2e}
\bibliography{refs}

\end{document}